\documentclass[]{interact}

\usepackage{epstopdf}
\usepackage{subcaption}
\usepackage{float}

\usepackage[longnamesfirst,sort]{natbib}
\bibpunct[, ]{(}{)}{;}{a}{,}{,}
\renewcommand\bibfont{\fontsize{10}{12}\selectfont}

\theoremstyle{plain}

\theoremstyle{definition}

\theoremstyle{remark}

\begin{document}

\articletype{}

\title{Fuzzy-based Higher Adaptive Order Sliding Mode Observers }

\author{
\name{Belhadjoudja Mohamed Camil}
\affil{National Polytechnic School, Department of Control Engineering, Algeria}
}

\maketitle

\begin{abstract}
In this article, I introduce the notion of Fuzzy-based Higher adaptive order sliding mode observers through the example of the super-twisting adaptive order sliding mode observer. I begin by presenting the super-twisting second order sliding mode observer for systems in triangular form. Then I replace the sign functions with fuzzy inference systems in order to eliminate the chattering effect. I then show that this variant of the super-twisting second order sliding mode observer may not converge depending on the input, so I replace the square root with an adaptive power $\gamma$. The efficiency of the proposed method is illustrated by simulations made on MATLAB/Simulink.
\end{abstract}


\section{Introduction}
The objective of this article is to introduce the notion of Fuzzy-based higher adaptive order sliding mode observers. I start by presenting the super-twisting second order sliding mode observer built in \citep{maamar}. This observer is presented in a version which makes it possible to estimate the states of a triangular system in the presence of delay in the output of the system. In this article, I do not consider the problem of delay in the output of the dynamical system. The super-twisting second order sliding mode observer, as other higher order sliding mode observers such as the third order sliding mode observer \citep{third}, is designed to reduce the chattering effect which is due to the discontinuity of the sign function. But as it has been shown in \citep{chattering}, even if the chattering effect decreases it may not go away. I show through simulations that the chattering effect is present in the super-twisting second order sliding mode observer presented in \citep{maamar}. In order to eliminate the chattering effect, I use a technique which has been presented in some articles \citep{fuzzy1,fuzzy2} and which consists of replacing sign functions with fuzzy inference systems. In addition, I show how an unknown input of a system in triangular form can be considered as an additional state to be estimated using a state observer such as the super-twisting second order sliding mode observer. As we will see in the simulations, the observer obtained using this technique no longer exhibits the chattering effect but may not converge depending on the input. In order to remedy this, I replace the square root of the super-twisting second order sliding mode observer with an adaptive power $\gamma$.

\section{The super-twisting second order sliding mode observer}
\begin{figure}
\begin{subfigure}{.5\textwidth}
  \centering
  \includegraphics[width=1\linewidth]{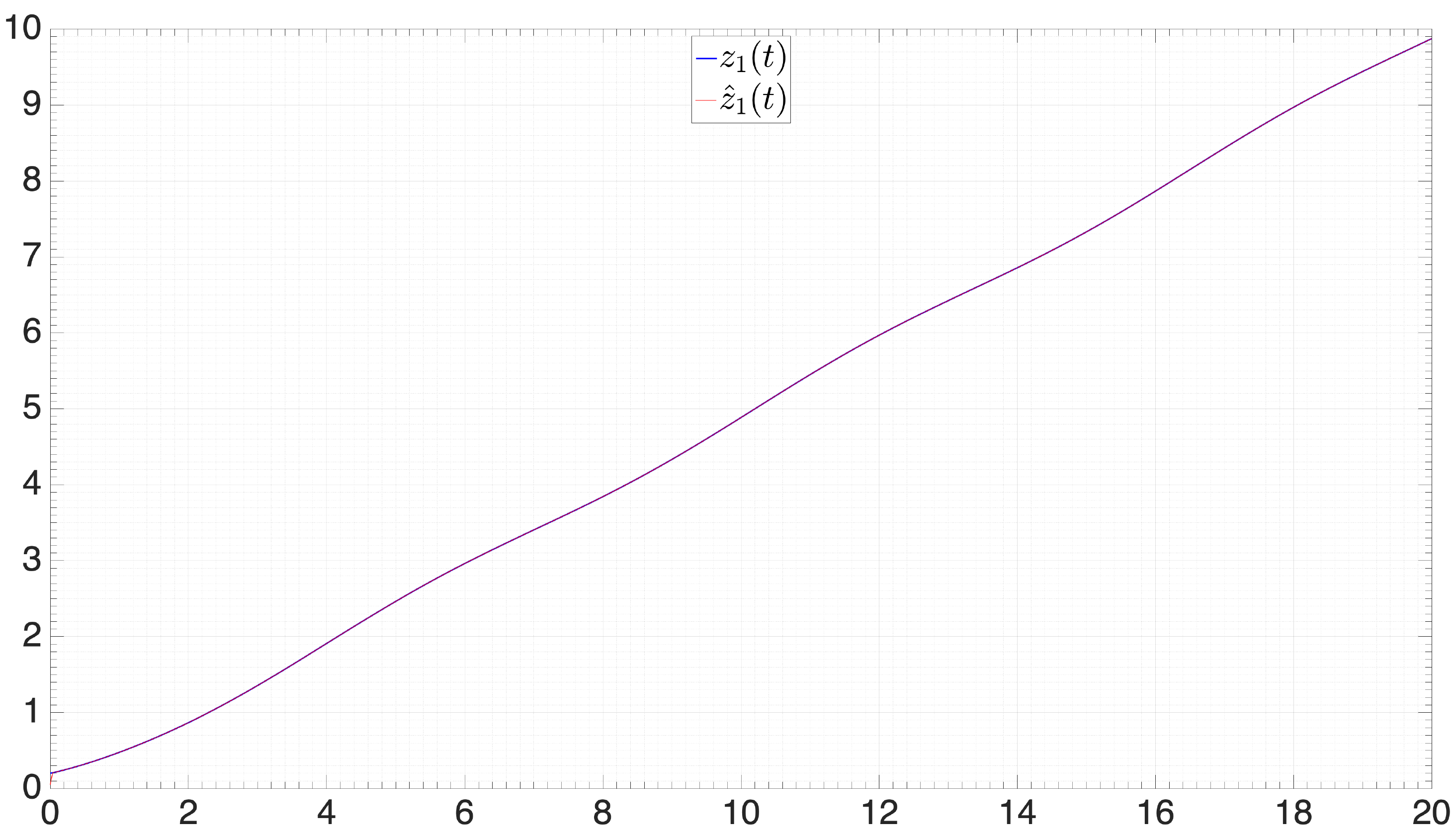}
  \caption{}
  \label{a}
\end{subfigure}
\begin{subfigure}{.5\textwidth}
  \centering
  \includegraphics[width=1\linewidth]{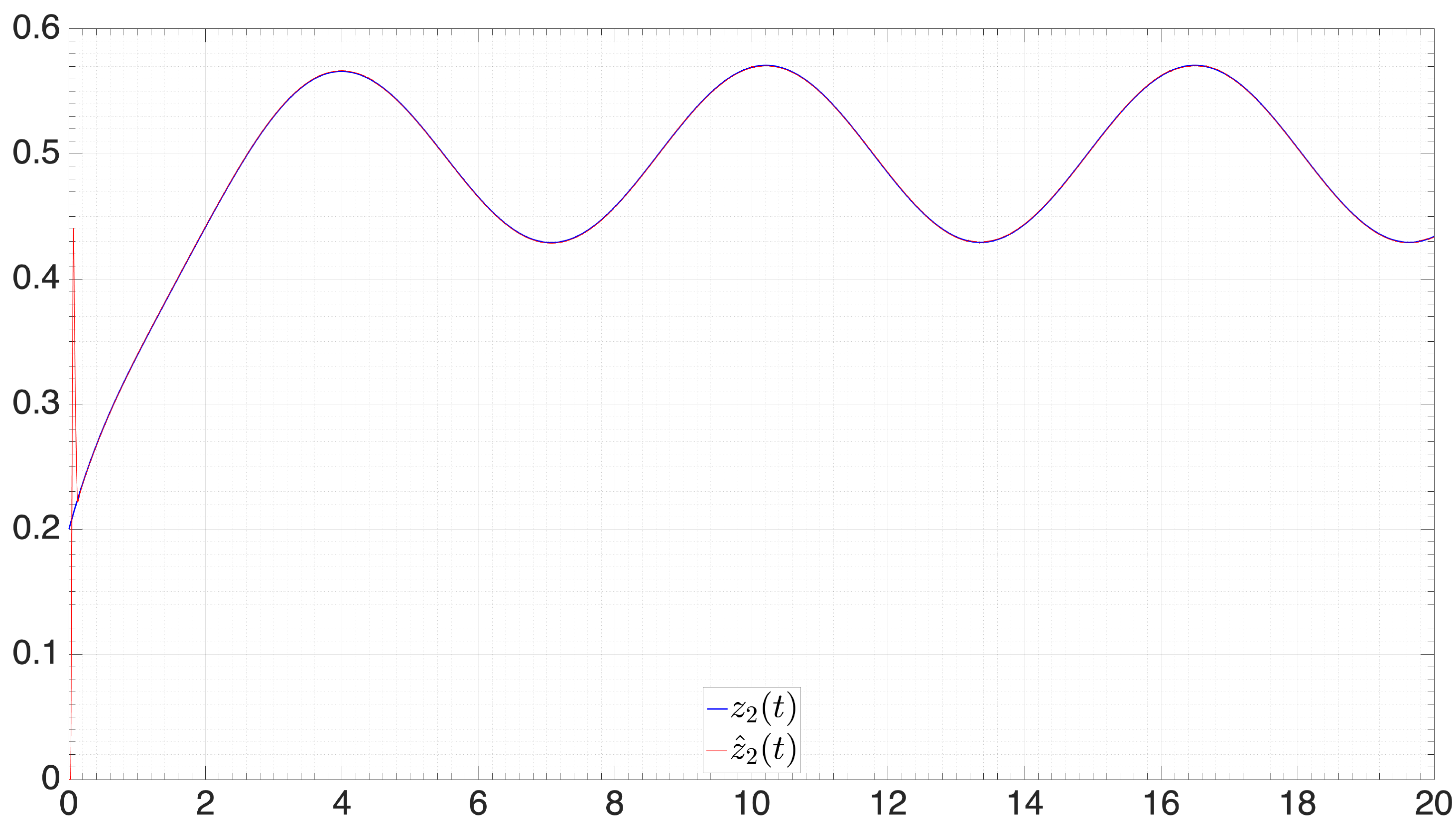}  
  \caption{}
  \label{b}
\end{subfigure}
\newline
\begin{subfigure}{.5\textwidth}
  \centering
  \includegraphics[width=1\linewidth]{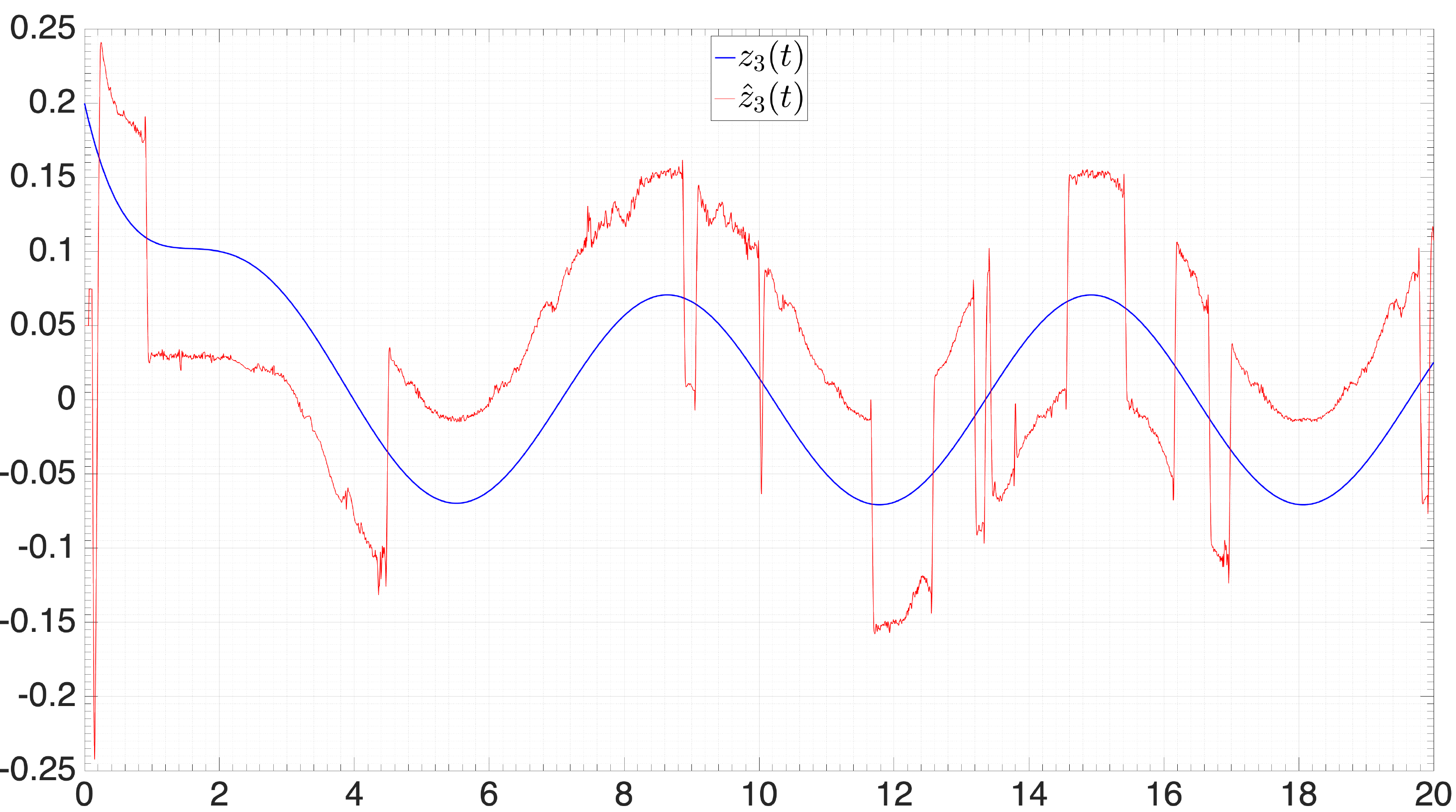}  
  \caption{}
  \label{c}
\end{subfigure}
\begin{subfigure}{.5\textwidth}
  \centering
  \includegraphics[width=1\linewidth]{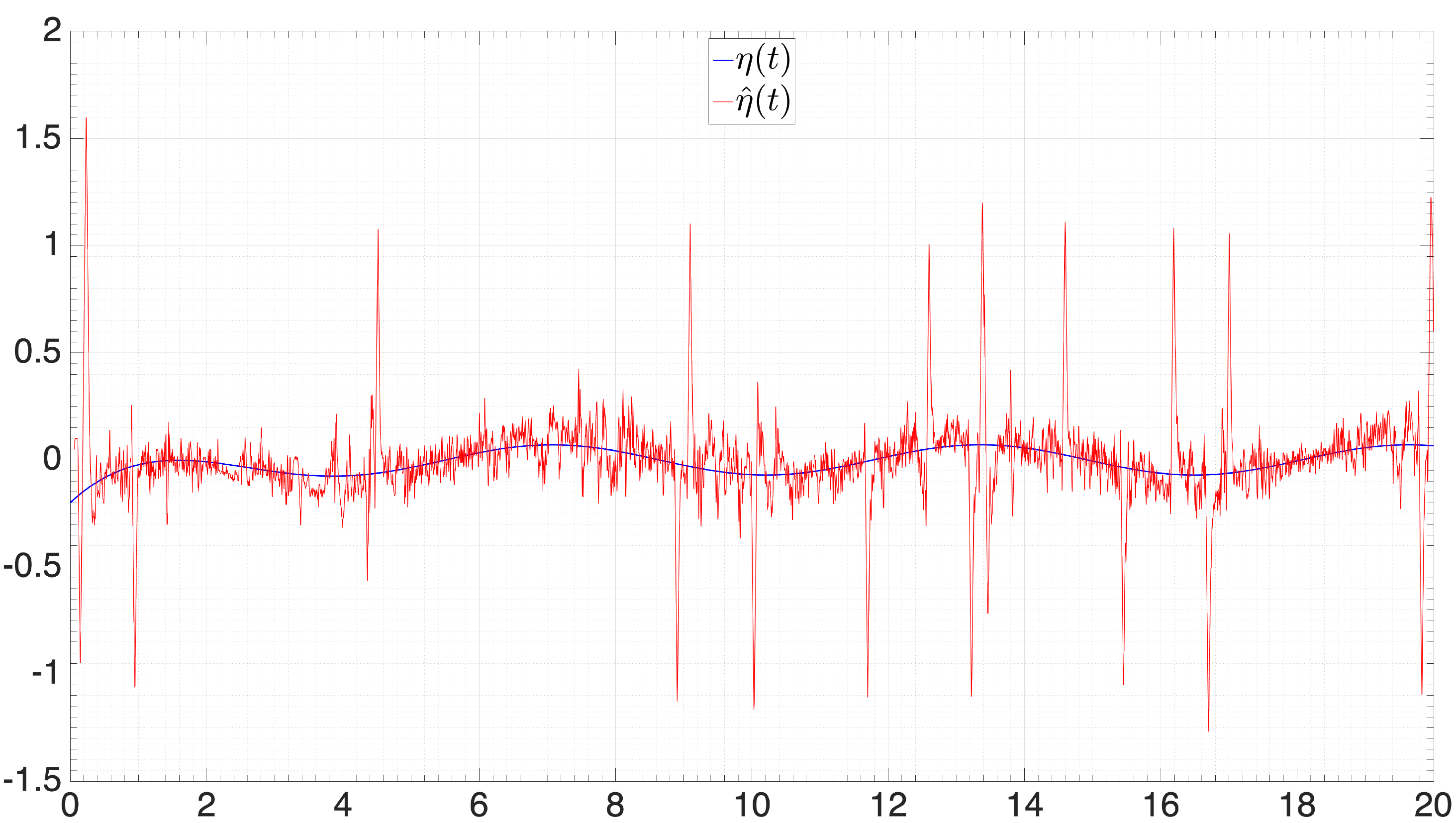}  
  \caption{}
  \label{d}
\end{subfigure}
\caption{The real signal in blue and the estimation in red}
\label{sim1}
\end{figure}

The super-twisting second order sliding mode observer is for systems of the form (\ref{triangular}).
\begin{equation} \label{triangular}
\left\lbrace
\begin{aligned}
&\dot{z}_{1}(t) = z_{2}(t) + \Phi_{1}(z_{1}) \\
&\dot{z}_{2}(t) = z_{3}(t) + \Phi_{2}(z_{1},z_{2}) \\
&... \\
&\dot{z}_{n-1}(t) = z_{n}(t) + \Phi_{n-1}(z_{1},z_{2},...,z_{n-1}) \\
&\dot{z}_{n}(t) = \eta(t) \\
&y(t) = z_{1}(t)
\end{aligned}
\right.
\end{equation}
where $\begin{bmatrix}z_{1} & z_{2} & ... & z_{n} \end{bmatrix}^T\in \mathbf{R}^n$ is the state vector, $y\in \mathbf{R}$ is the output, $\eta \in \mathbf{R}$ and $\Phi_{i}$, for $i\in \lbrace 1,2,...,n-1\rbrace$, is a nonlinear function. 
\\ \\ The super-twisting second order sliding mode observer for (\ref{triangular}), and constructed in \citep{maamar}, is given by (\ref{triangularObserver}).   
\begin{equation} \label{triangularObserver}
\left \lbrace
\begin{aligned}
&\dot{\hat{z}}_{1} = \Tilde{z}_{2} + \lambda_{1}\sqrt{|\epsilon _{1}|} sign(\epsilon _{1}) + \Phi_{1}(z_{1})  \\
&\dot{\Tilde{z}}_{2} = \alpha_{1}sign(\epsilon _{1}) \\
&\dot{\hat{z}}_{2} = E_{1}[\Tilde{z}_{3} + \lambda_{2}\sqrt{|\epsilon _{2}|} sign(\epsilon _{2}) + \Phi_{2}(z_{1}, \Tilde{z}_{2})] \\
&\dot{\Tilde{z}}_{3} = E_{1}[\alpha_{2}sign(\epsilon _{2})]  \\
&\dot{\hat{z}}_{3} = E_{2}[\Tilde{z}_{4} + \lambda_{3}\sqrt{|\epsilon _{3}|} sign(\epsilon _{3}) + \Phi_{3}(z_{1}, \Tilde{z}_{2}, \Tilde{z}_{3})]  \\
&...  \\
&\dot{\Tilde{z}}_{n-1} = E_{n-3}[\alpha_{n-2}sign(\epsilon _{n-2})] \\
&\dot{\hat{z}}_{n-1} = E_{n-2}[\Tilde{z}_{n} + \lambda_{n-1}\sqrt{|\epsilon _{n-1}|} sign(\epsilon _{n-1}) + \Phi_{n-1}(z_{1}, \Tilde{z}_{2}, ..., \Tilde{z}_{n-1})]  \\
&\dot{\Tilde{z}}_{n} = E_{n-2}[\alpha_{n-1}sign(\epsilon _{n-1})]  \\
&\dot{\hat{z}}_{n} = E_{n-1}[\Tilde{\theta} + \lambda_{n}\sqrt{|\epsilon _{n}|} sign(\epsilon _{n})]  \\
&\dot{\Tilde{\theta}} = E_{n-1}[\alpha_{n}sign(\epsilon_{n} )] 
\end{aligned}
\right.
\end{equation}
where :
\begin{equation}
\begin{aligned}
&\epsilon_{i}  = \Tilde{z}_{i}  - \hat{z}_{i}  \nonumber \\
&\Tilde{z}_{1}  = z_{1}  = y  \Longrightarrow \epsilon_{1}  = e_{1}  = z_{1}  - \hat{z}_{i}  \nonumber
\end{aligned}
\end{equation}
and $\hat{z}_{i} $ and $\Tilde{z}_{i} $ are the estimated state and the internal state of the observer, respectively. The function $E_{i}$ is given by (\ref{Eifunction}). 
\begin{equation} \label{Eifunction}
E_{i} = 
\left\lbrace
\begin{aligned}
&1 \ \ \ \ \ if \ \ |\epsilon_{j}  |\leq \epsilon, \ \ \forall j\leq i\\
&0 \ \ \ \ \ else
\end{aligned}
\right.
\end{equation}
where $\epsilon $ is a small positive constant. \\ \\ 
The parameters $\alpha_{i}$ are the observer gains and the parameters $\lambda_{i}>0$ are the correction factors that are here to ensure the convergence of the observer. 
\\ \\ Under some assumptions \citep{maamar}, the observer (\ref{triangularObserver}) converges. There is, however, the chattering effect as we will see in the following example. 
\\ \\ Let us consider the system in triangular form given by (\ref{simpledynamics}). 
\begin{equation} \label{simpledynamics}
\left\lbrace
\begin{aligned}
&\dot{z}_{1} = z_{2}\\
&\dot{z}_{2} = z_{3}\\
&\dot{z}_{3} = -z_{3} + d = \eta (t)
\end{aligned}
\right.
\end{equation}
where $d=0.1sin(t)$ is an external input. We take $\epsilon = 0.025$, and the gains:
\begin{equation}
\begin{aligned}
&\alpha_{1} = 30; \ \ \alpha_{2} = 30; \ \ \alpha_{3} = 30; \nonumber \\
&\lambda_{1} = 15; \ \ \lambda_{2} = 15; \ \ \lambda_{2} = 15. 
\end{aligned}
\end{equation}
and the initial conditions: 
\begin{equation}
\begin{aligned}
&\begin{bmatrix} z_{1}(0) & z_{2}(0) & z_{3}(0) \end{bmatrix}^T = \begin{bmatrix} 0.2 & 0.2 & 0.2 \end{bmatrix}^T  \nonumber \\
&\begin{bmatrix} \hat{z}_{1}(0) & \hat{z}_{2}(0) & \hat{z}_{3}(0) \end{bmatrix}^T = \begin{bmatrix} 0.05 & 0 & 0.05 \end{bmatrix}^T \\
&\begin{bmatrix} \Tilde{z}_{2}(0) & \Tilde{z}_{3}(0) & \ \Tilde{\theta}(0) \end{bmatrix}^T = \begin{bmatrix} 0.05 & 0 & 0.05 \end{bmatrix}^T
\end{aligned}
\end{equation}
I did the simulations on MATLAB/Simulink, with a time step of $10^{-5}$. The results are shown in Figure \ref{sim1}.
As we can see, there is the chattering effect, which is due to the discontinuity of the sign function. In order to eliminate this effect, a technique has been proposed in some articles \citep{fuzzy1,fuzzy2} which consists in replacing the sign functions by fuzzy inference systems. I will apply this technique on the super twisting second order sliding mode observer.
\section{Fuzzy-based super-twisting second order sliding mode observer for state and unknown input estimation}
\begin{figure}
    \centering
    \includegraphics[width=1\linewidth]{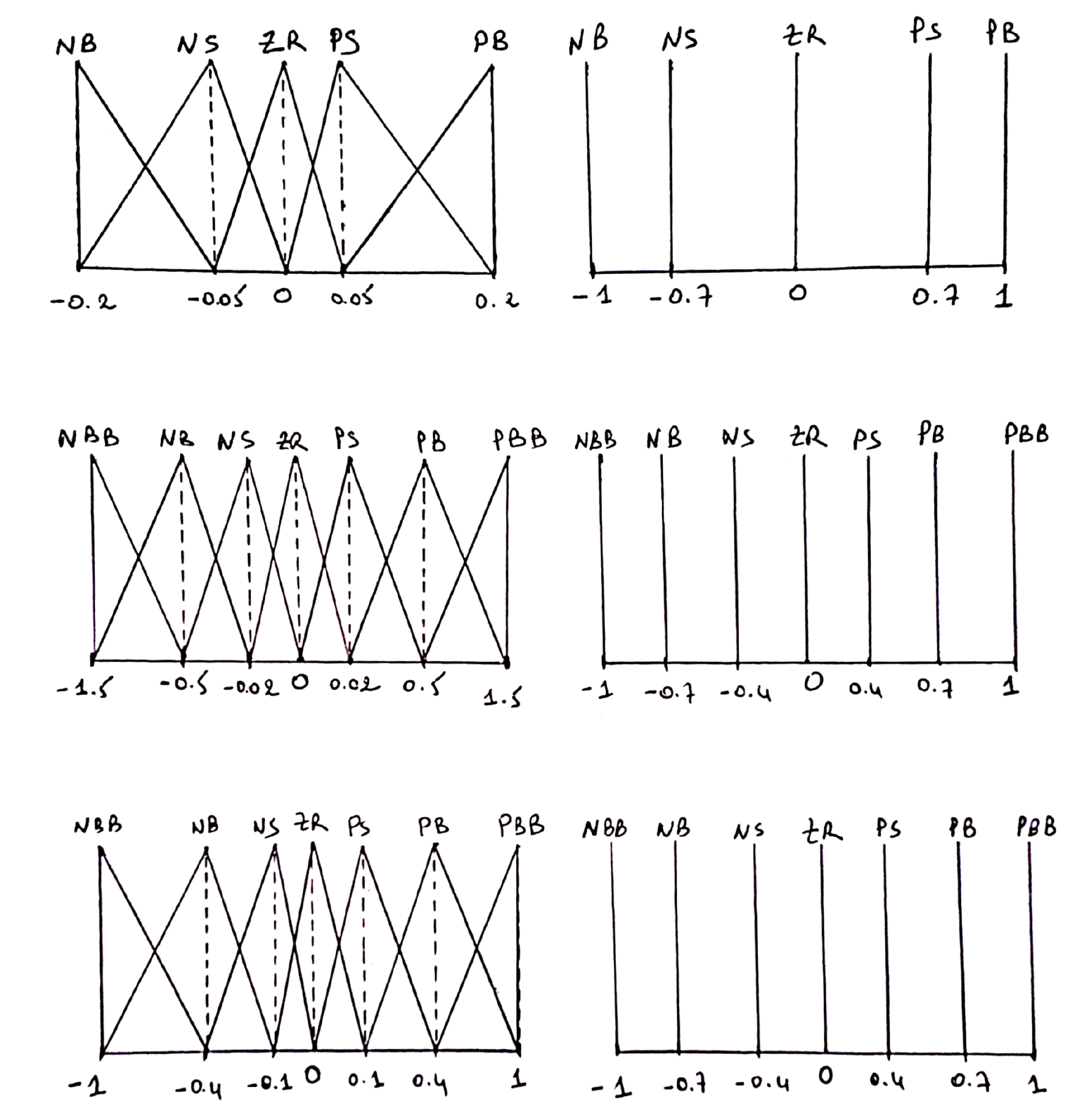}
    \caption{The membership functions (on the left the input and on the right the output)}
    \label{fuzzysystems}
\end{figure}

A first remark is that for a system in triangular form, it is possible to consider an unknown input as being an additional state of the system, and thus use a state observer to estimate the unknown input. For example, for the dynamical system (\ref{simpledynamics}), it suffices to set $z_{4} = d$ in order to obtain the new system (\ref{newsimpledynamics}) and thus estimate $d$ using the super-twisting second order sliding mode observer.
\begin{equation} \label{newsimpledynamics}
\left\lbrace
\begin{aligned}
&\dot{z}_{1} = z_{2}\\
&\dot{z}_{2} = z_{3}\\
&\dot{z}_{3} = -z_{3} + z_{4}\\
&\dot{z}_{4} = \dot{d} = \eta (t)
\end{aligned}
\right.
\end{equation}
In order to eliminate the chattering effect, I replace the sign functions with fuzzy inference systems as it has been done in \citep{fuzzy1,fuzzy2}. For the observer (\ref{newsimpledynamics}), there are four sign functions. I design three fuzzy inference systems: $\psi_{1}$, $\psi_{2}$ and $\psi_{3}$ whose membership functions are shown in Figure \ref{fuzzysystems} (line $i$ corresponds to $\psi_{i}$). The rules are of the form: 
\\ \\ - If $\epsilon_{i} $ is \textbf{NBB} then $\psi_{i}$ is \textbf{NBB}. \\
- If $\epsilon_{i} $ is \textbf{NB} then $\psi_{i}$ is \textbf{NB}. \\
- If $\epsilon_{i} $ is \textbf{NS} then $\psi_{i}$ is \textbf{NS}. \\
- If $\epsilon_{i} $ is \textbf{ZR} then $\psi_{i}$ is \textbf{ZR}. \\
- If $\epsilon_{i} $ is \textbf{PS} then $\psi_{i}$ is \textbf{PS}. \\
- If $\epsilon_{i} $ is \textbf{PB} then $\psi_{i}$ is \textbf{PB}. \\
- If $\epsilon_{i} $ is \textbf{PBB} then $\psi_{i}$ is \textbf{PBB}.\\ \\
for $i=2,3$ (for $\epsilon_{4}$, I will use $\psi_{3}$). For $i=1$, the rules are: \\ \\
- If $\epsilon_{1} $ is \textbf{NB} then $\psi_{1}$ is \textbf{NB}. \\
- If $\epsilon_{1} $ is \textbf{NS} then $\psi_{1}$ is \textbf{NS}. \\
- If $\epsilon_{1} $ is \textbf{ZR} then $\psi_{1}$ is \textbf{ZR}. \\
- If $\epsilon_{1} $ is \textbf{PS} then $\psi_{1}$ is \textbf{PS}. \\
- If $\epsilon_{1} $ is \textbf{PB} then $\psi_{1}$ is \textbf{PB}. \\ \\
where: \\ \\
- \textbf{NBB} : Negative Big Big.\\
- \textbf{NB} : Negative Big.\\
- \textbf{NS} : Negative Small.\\
- \textbf{ZR} : Zero.\\
- \textbf{PS} : Positive Small.\\
- \textbf{PBB} : Positive Big Big.\\
The fuzzy-based super-twisting second order sliding mode observer for (\ref{newsimpledynamics}) is given by (\ref{fuzzysuper}).
\begin{equation} \label{fuzzysuper}
\left \lbrace
\begin{aligned}
&\dot{\hat{z}}_{1} = \Tilde{z}_{2} + \lambda_{1}\sqrt{|\epsilon _{1}|} \psi_{1}(\epsilon_{1})  \\
&\dot{\Tilde{z}}_{2} = \alpha_{1}\psi_{1}(\epsilon_{1}) \\
&\dot{\hat{z}}_{2} = E_{1}[\Tilde{z}_{3} + \lambda_{2}\sqrt{|\epsilon _{2}|} \psi_{2}(\epsilon_{2})] \\
&\dot{\Tilde{z}}_{3} = E_{1}[\alpha_{2}\psi_{2}(\epsilon_{2})]  \\
&\dot{\hat{z}}_{3} = E_{2}[\Tilde{z}_{4} + \lambda_{3}\sqrt{|\epsilon _{3}|} \psi_{3}(\epsilon_{3}) - \Tilde{z}_{3}]  \\
&\dot{\Tilde{z}}_{4} = E_{2}[\alpha_{3}\psi_{3}(\epsilon_{3})]  \\
&\dot{\hat{z}}_{4} = E_{3}[\Tilde{\theta} + \lambda_{4}\sqrt{|\epsilon _{4}|} \psi_{3}(\epsilon_{4})]  \\
&\dot{\Tilde{\theta}} = E_{3}[\alpha_{4}\psi_{3}(\epsilon_{4})] 
\end{aligned}
\right.
\end{equation}
where $\lambda_{4}=15$, $\alpha_{4}=30$, and the initial conditions are: 
\begin{equation}
\begin{aligned}
&\begin{bmatrix} z_{1}(0) & z_{2}(0) & z_{3}(0) & z_{4}(0)\end{bmatrix}^T = \begin{bmatrix} 0.2 & 0.2 & 0.2 & 0.2\end{bmatrix}^T  \nonumber \\
&\begin{bmatrix} \hat{z}_{1}(0) & \hat{z}_{2}(0) & \hat{z}_{3}(0) & \hat{z}_{4}(0) \end{bmatrix}^T = \begin{bmatrix} 0.05 & 0 & 0.05 & 0.05 \end{bmatrix}^T \\
&\begin{bmatrix} \Tilde{z}_{2}(0) & \Tilde{z}_{3}(0) & \Tilde{z}_{4}(0) & \ \Tilde{\theta}(0) \end{bmatrix}^T = \begin{bmatrix} 0.05 & 0 & 0 & 0.05 \end{bmatrix}^T
\end{aligned}
\end{equation}
The other constants are the same as for the precedent simulations.
\begin{figure}[H]
\begin{subfigure}{.5\textwidth}
  \centering
  \includegraphics[width=1\linewidth]{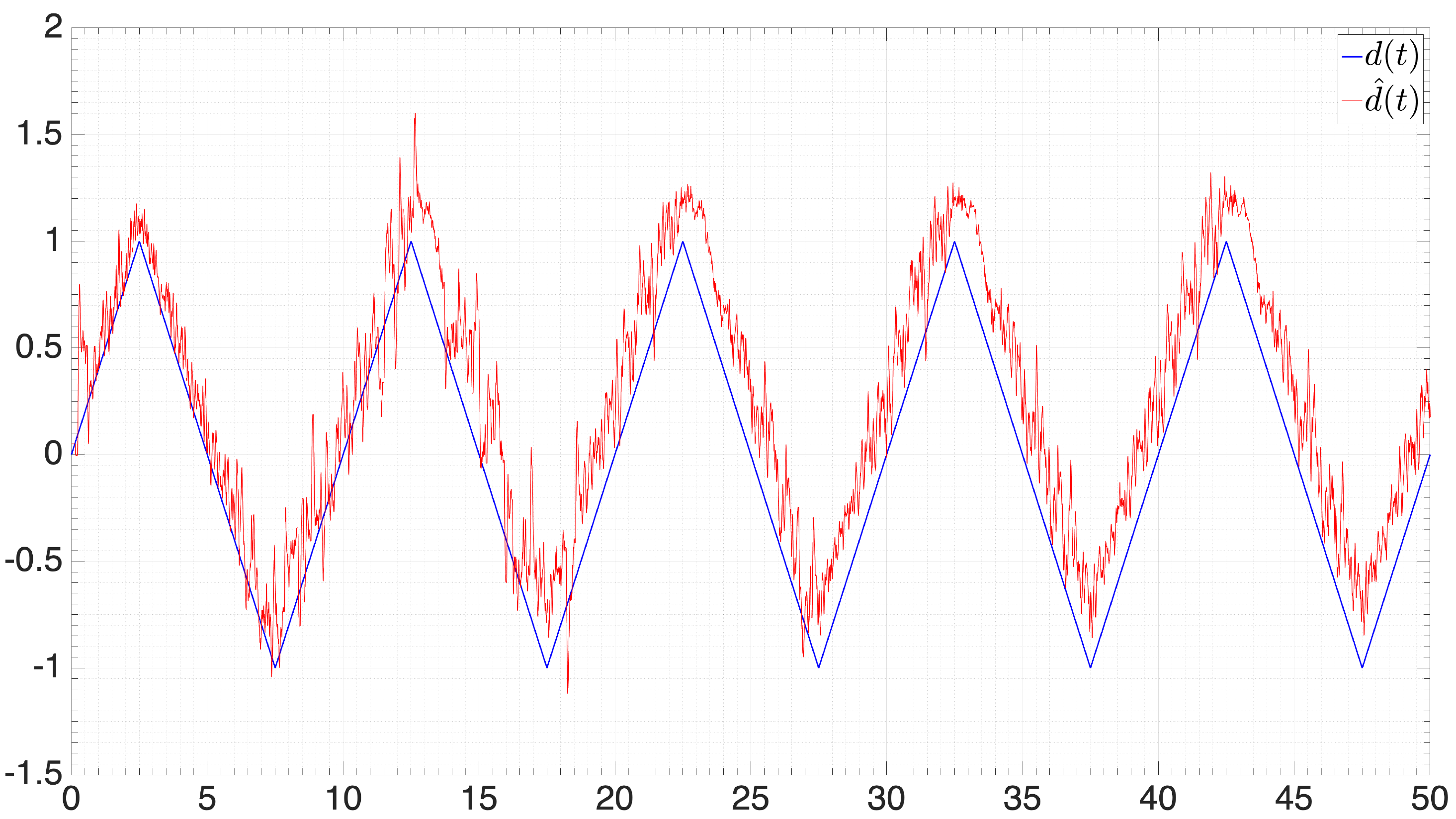}
  \caption{}
  \label{a2}
\end{subfigure}
\begin{subfigure}{.5\textwidth}
  \centering
  \includegraphics[width=1\linewidth]{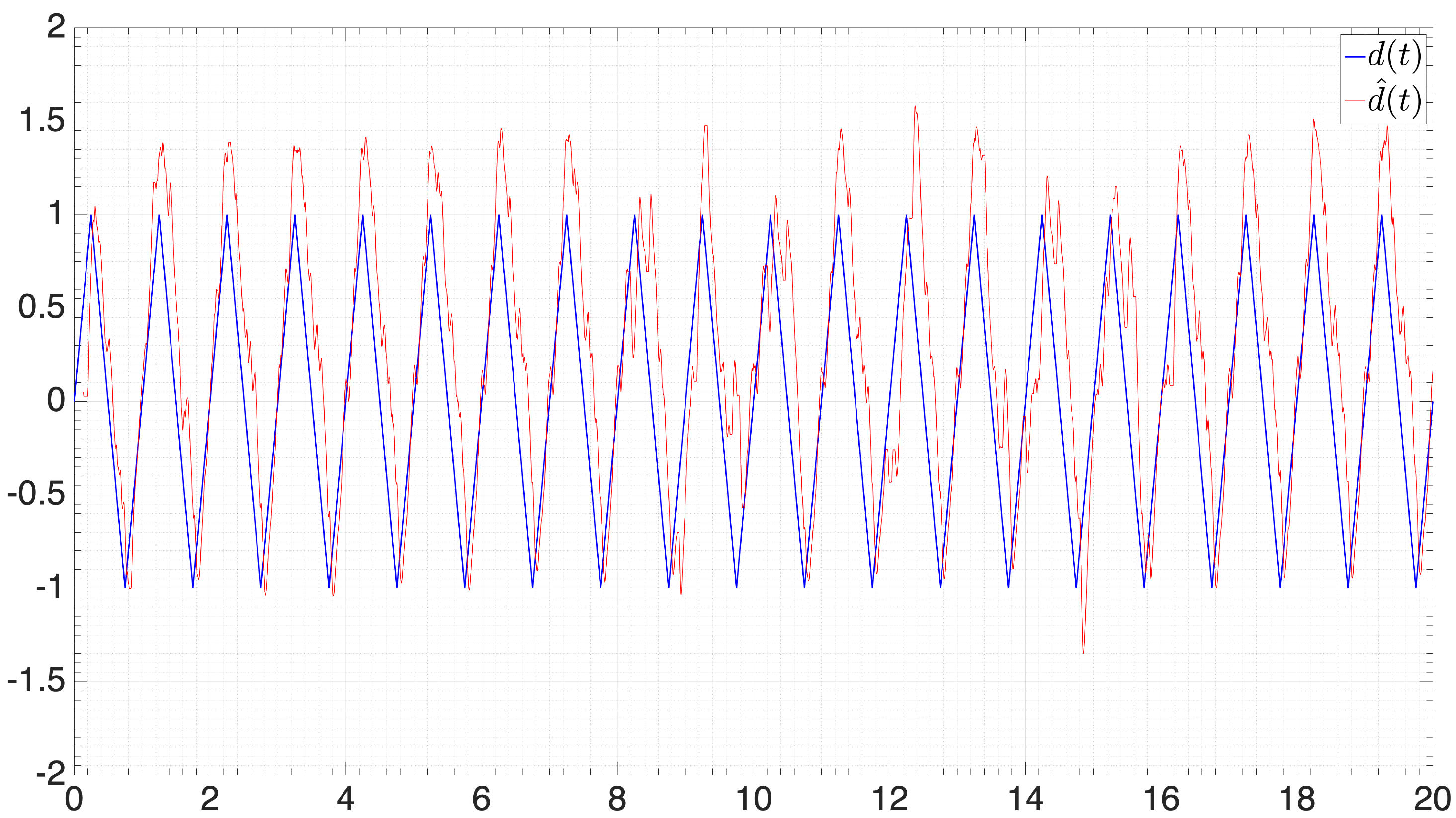}  
  \caption{}
  \label{b2}
\end{subfigure}
\newline
\begin{subfigure}{.5\textwidth}
  \centering
  \includegraphics[width=1\linewidth]{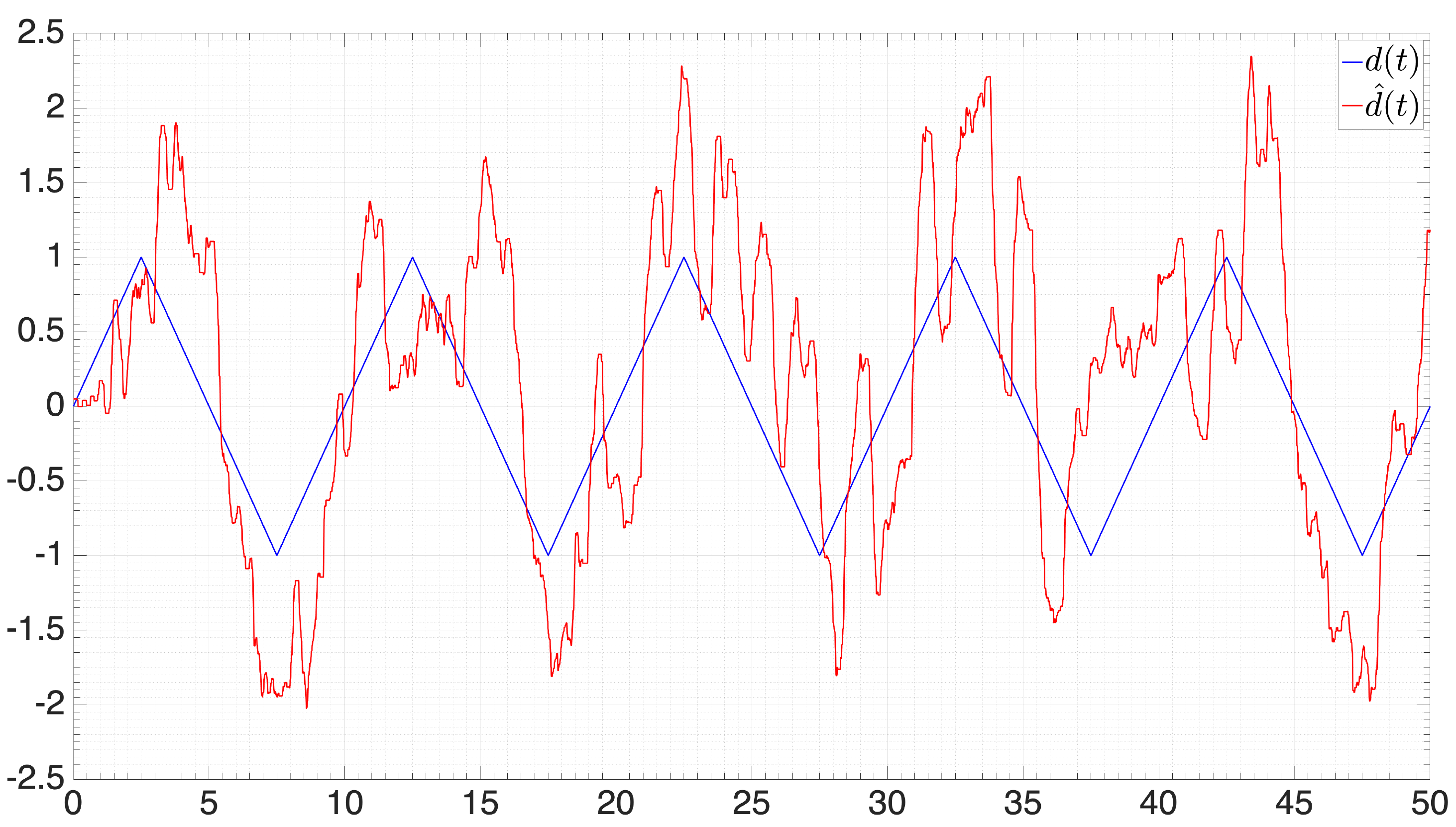}  
  \caption{}
  \label{c2}
\end{subfigure}
\begin{subfigure}{.5\textwidth}
  \centering
  \includegraphics[width=1\linewidth]{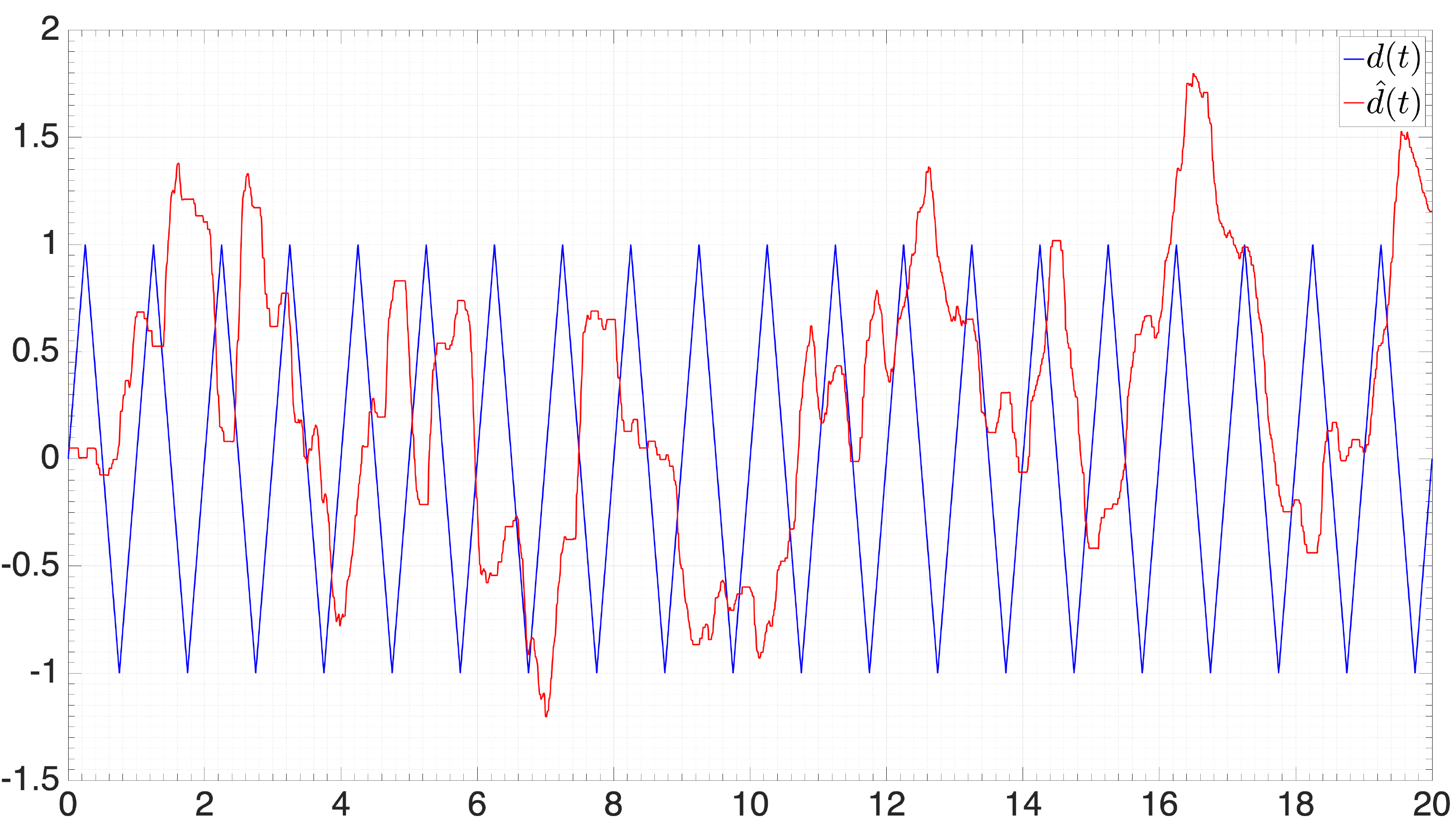}  
  \caption{}
  \label{d2}
\end{subfigure}
\newline
\begin{subfigure}{.5\textwidth}
  \centering
  \includegraphics[width=1\linewidth]{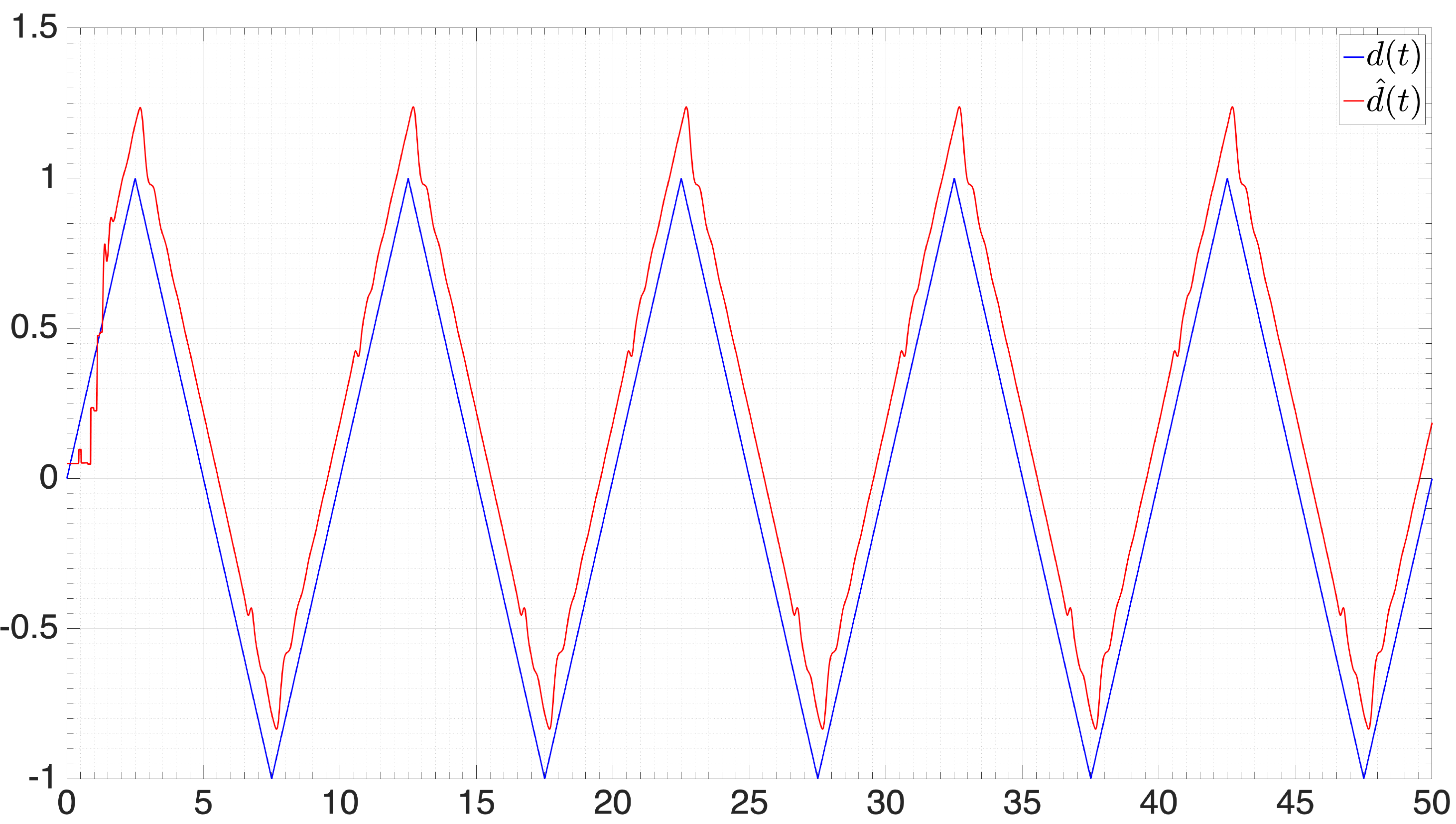}  
  \caption{}
  \label{e2}
\end{subfigure}
\begin{subfigure}{.5\textwidth}
  \centering
  \includegraphics[width=1\linewidth]{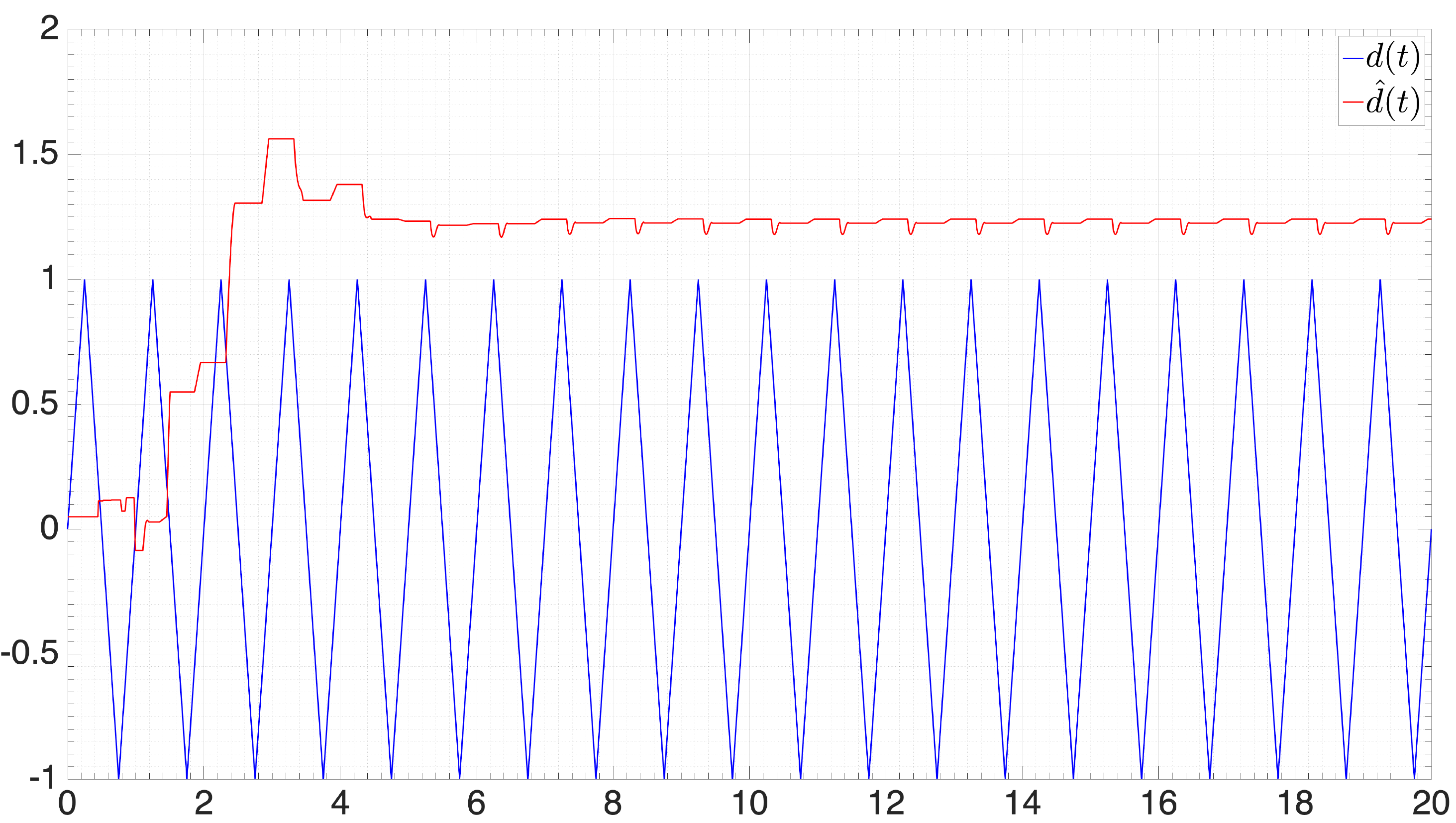}  
  \caption{}
  \label{f2}
\end{subfigure}
\caption{Fisrt line: using sign, time step $= 10^{-5}$; Second line: using sign, time step $= 10^{-3}$; Third line: Fuzzy, time step $= 10^{-3}$}
\label{simz}
\end{figure}
The results of the simulations are shown in Figure \ref{simz}. As we can see, using fuzzy inference systems eliminates the chattering effect. However, for a higher frequency input, we notice that the estimation by fuzzy inference systems is worse than the one using the signs functions. So the problem is that fuzzy inference systems can eliminate the chattering effect, but depending on the input, fuzzy inference systems can produce worse estimates than those obtained using signs functions. To solve this problem, I introduce the notion of Fuzzy-based Higher adaptive order sliding mode observers.
\section{The Fuzzy-based Super-twisting adaptive order sliding mode observer}
\begin{figure}
\begin{subfigure}{.5\textwidth}
  \centering
  \includegraphics[width=1\linewidth]{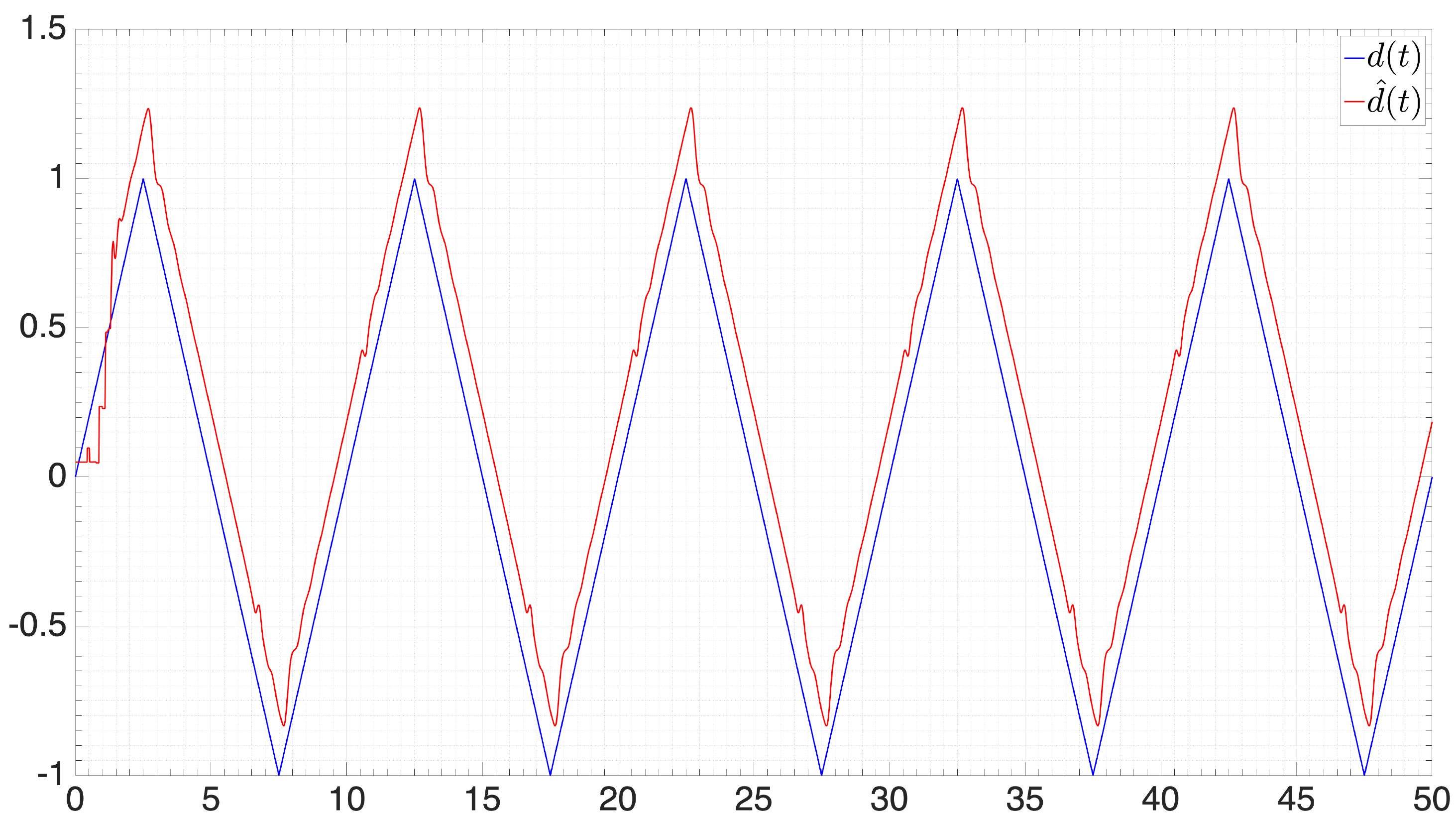}
  \caption{}
  \label{a3}
\end{subfigure}
\begin{subfigure}{.5\textwidth}
  \centering
  \includegraphics[width=1\linewidth]{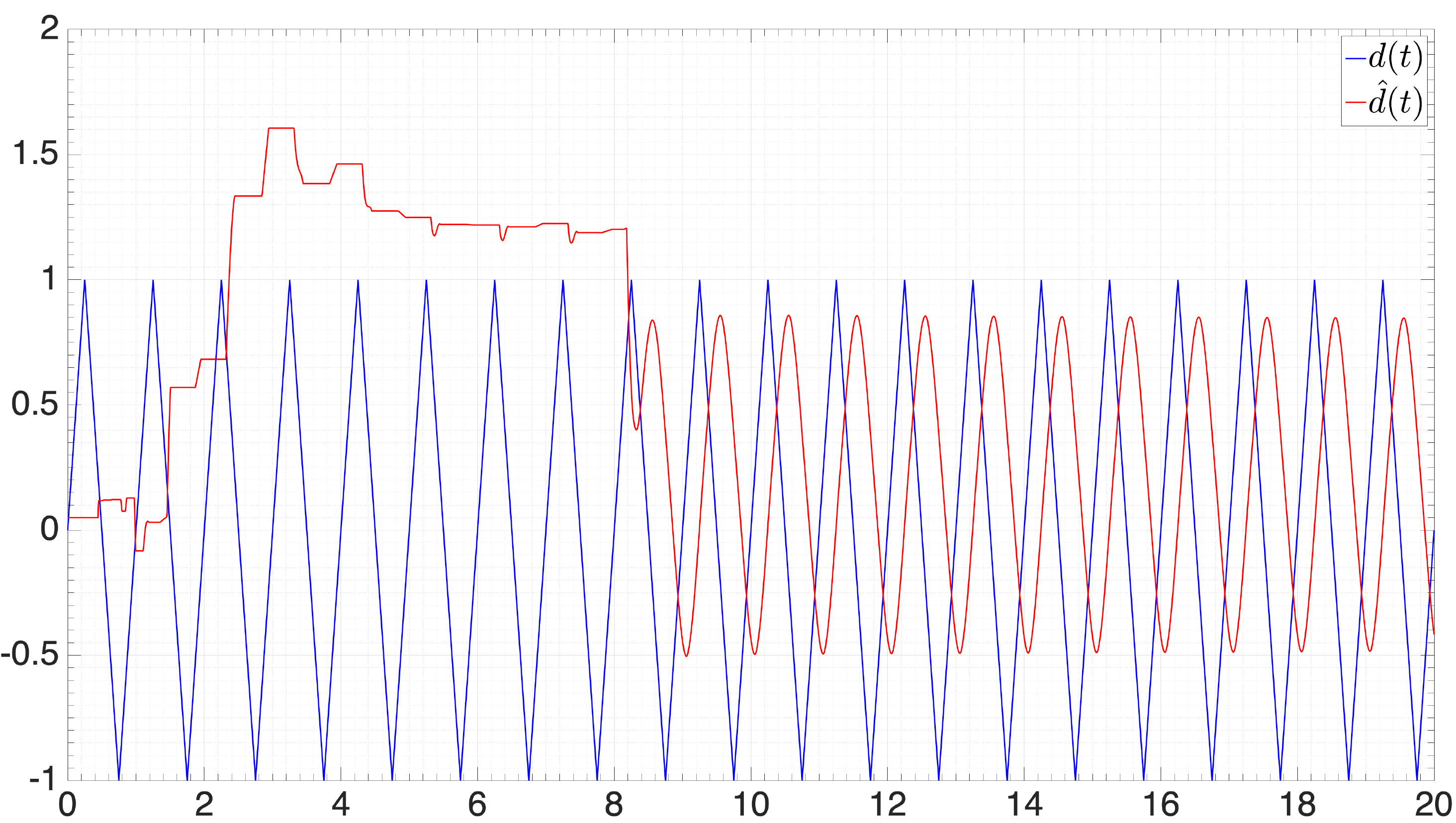}  
  \caption{}
  \label{b3}
\end{subfigure}
\newline
\begin{subfigure}{.5\textwidth}
  \centering
  \includegraphics[width=1\linewidth]{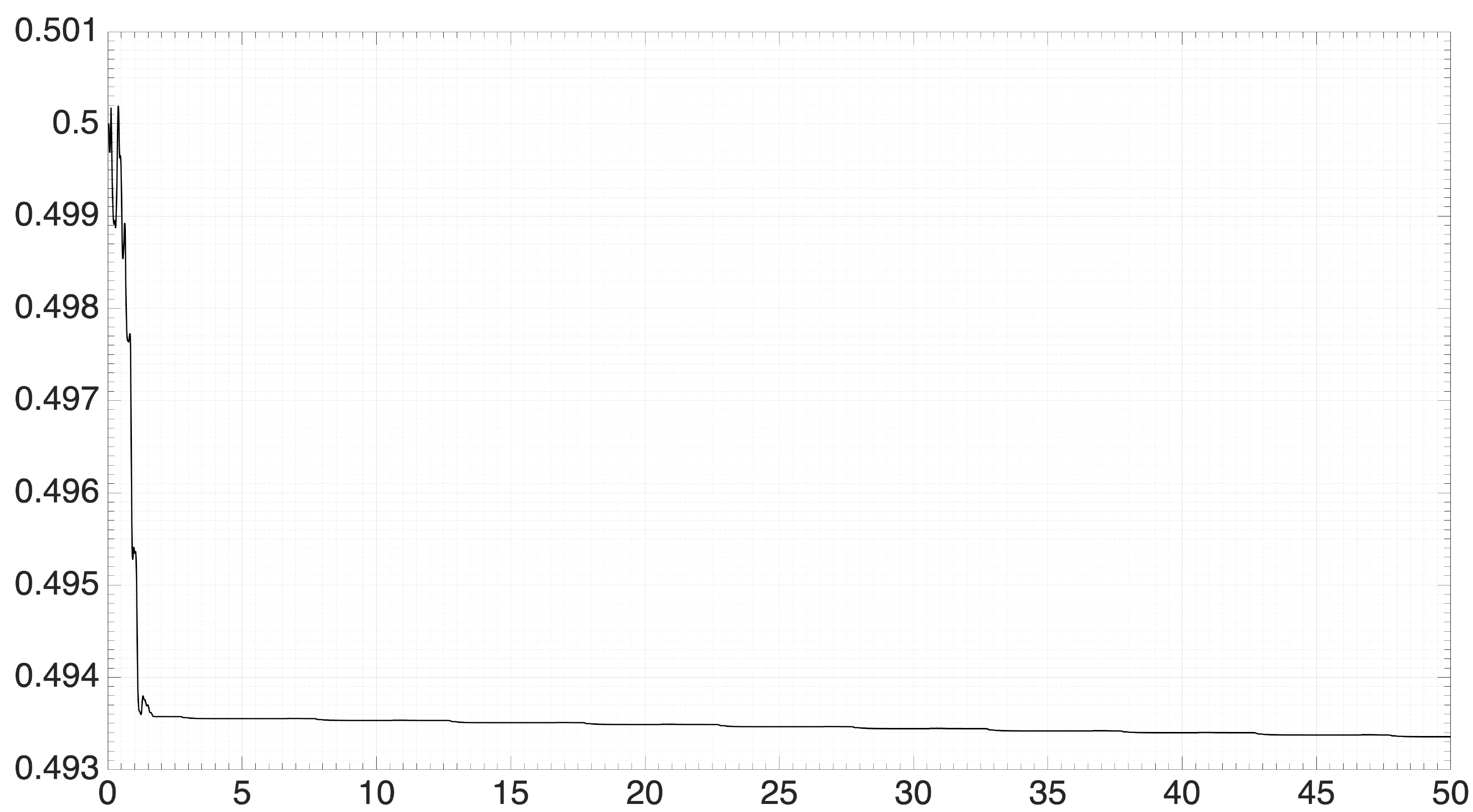}  
  \caption{}
  \label{c3}
\end{subfigure}
\begin{subfigure}{.5\textwidth}
  \centering
  \includegraphics[width=1\linewidth]{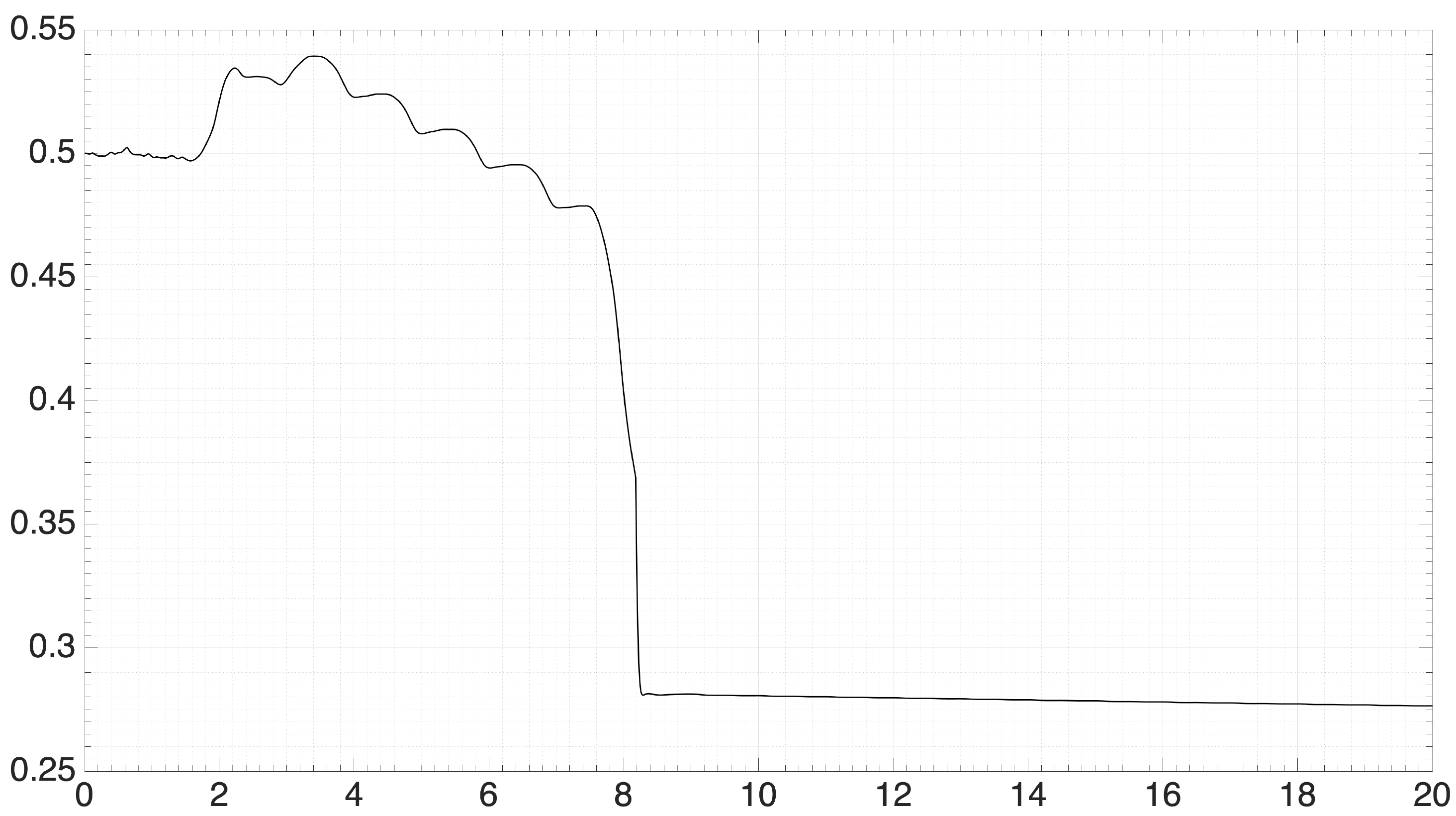}  
  \caption{}
  \label{d3}
\end{subfigure}
\caption{Top: the estimation; Bottom: the corresponding $\gamma$}
\label{simu}
\end{figure}

The structure of the Fuzzy-based Super-twisting adaptive order sliding mode observer is given by (\ref{adaptivesupertwistingobserverfuzzymodified}). 
\begin{equation} \label{adaptivesupertwistingobserverfuzzymodified}
\left\lbrace
\begin{aligned}
&\dot{\hat{z}}_{1}  = \Tilde{z}_{2}  + \lambda_{1}|\epsilon _{1} |^\gamma \psi_{1}(\epsilon _{1} ) + \Phi_{1}(z_{1} )  \\
&\dot{\Tilde{z}}_{2}  = \alpha_{1}\psi_{1}(\epsilon _{1} ) \\
&\dot{\hat{z}}_{2}  = E_{1}[\Tilde{z}_{3}  + \lambda_{2}|\epsilon _{2} |^\gamma \psi_{2}(\epsilon _{2} ) + \Phi_{2}(z_{1} , \Tilde{z}_{2} )] \\
&\dot{\Tilde{z}}_{3}  = E_{1}[\alpha_{2}\psi_{2}(\epsilon _{2} )]  \\
&\dot{\hat{z}}_{3}  = E_{2}[\Tilde{z}_{4}  + \lambda_{3}|\epsilon _{3} |^\gamma \psi_{3}(\epsilon _{3} ) + \Phi_{3}(z_{1} , \Tilde{z}_{2} , \Tilde{z}_{3} )]  \\
&...  \\
&\dot{\Tilde{z}}_{n-1}  = E_{n-3}[\alpha_{n-2}\psi_{n-2}(\epsilon _{n-2} )] \\
&\dot{\hat{z}}_{n-1}  = E_{n-2}[\Tilde{z}_{n}  + \lambda_{n-1}|\epsilon _{n-1} |^\gamma \psi_{n-1}(\epsilon _{n-1} ) + \Phi_{n-1}(z_{1} , \Tilde{z}_{2} , ..., \Tilde{z}_{n-1}  )]  \\
&\dot{\Tilde{z}}_{n}\tau = E_{n-2}[\alpha_{n-1}\psi_{n-1}(\epsilon _{n-1} )]  \\
&\dot{\hat{z}}_{n}  = E_{n-1}[\Tilde{d}  + \lambda_{n}|\epsilon _{n} |^\gamma \psi_{n}(\epsilon _{n} ) + \Phi_{n}(z_{1} , \Tilde{z}_{2} , ..., \Tilde{z}_{n-1} , \Tilde{z}_{n}  )]  \\
&\dot{\Tilde{d}}\tau = E_{n-1}[\alpha_{n}\psi_{n}(\epsilon _{n} )]  \\
&\dot{\hat{d}}  = E_{n}[\Tilde{\theta}  + \lambda_{n+1}|\epsilon _{n+1} |^\gamma \psi_{n+1}(\epsilon _{n+1} )]  \\
&\dot{\Tilde{\theta}}  = E_{n}[\alpha_{n+1}\psi_{n+1}(\epsilon_{n+1}  )] 
\end{aligned}
\right.
\end{equation}
As we can see, the power $1/2$ is replaced by a function $\gamma$. By changing the value of $\gamma$, the convergence precision of the observer changes. We can try for example to minimize the errors $\epsilon_{j}$ using the adaptation law (\ref{adaptationlawgamma}). 
\begin{equation} \label{adaptationlawgamma}
\dot{\gamma} = - \Gamma \nabla_{\gamma}\mathcal{L}
\end{equation}
where $\Gamma$ is the learning rate and $\mathcal{L}$ is a loss function, in our case (\ref{lossfunction}). 
\begin{equation} \label{lossfunction}
\mathcal{L} = |\epsilon_{1} |^2 + |\epsilon_{2} |^2 + ... + |\epsilon_{n} |^2
\end{equation}
The results of the simulations on the precedent example for $\Gamma = 0.004$ and $\gamma (0)=0.5$ are shown in Figure \ref{simu} (the time step is set to $10^{-3}$).
As we can see, using an adaptive order improves the performance of the observer.
\section{Conclusion} 
In this article, I first showed through simulations that the super-twisting second order sliding mode observer presented in \citep{maamar} can produce high frequency oscillations in the estimate (the chattering effect) due to the discontinuity of the sign function. I then showed that one could consider an unknown input of a system in triangular form as being an additional state to be estimated using a state observer such as the super-twisting second order sliding mode observer. Then, I used a technique presented in some articles \citep{fuzzy1, fuzzy2} which consists in replacing the sign functions by fuzzy inference systems in order to eliminate the chattering effect. I showed through the simulations that this technique works on the super-twisting second order sliding mode observer but that depending on the input, the fuzzy-based observer could converge less well than the original observer with the sign functions. In order to remedy this, I introduced the notion of Fuzzy-based super-twisting adaptive order sliding mode observer which is a variant of the fuzzy-based super-twisting second order sliding mode observer where the power $1/2$ is replaced by an adaptive power $\gamma$. I showed through simulations the efficiency of the proposed method. This method can easily be generalized to other types of higher order sliding mode observers, giving rise to the notion of Fuzzy-based Higher Adaptive Order sliding mode observers.

\end{document}